# Microcanonical ensemble and incompleteness of quantum mechanics

V. A. Skrebnev

Physics Department, Kazan State University, Russian Federation

e-mail: vskrebnev@mail.ru

It is shown, that the only reason for possibility of using microcanonical ensemble is that there are probabilistic processes in microworld, that are not described by quantum mechanic. On a simple example it is demonstrated, that canonical distribution is not independent and equal to microcanonical, but is a result of averaging by microcanonical ensemble.

In quantum statistical mechanics only eigenstates of the system are considered to be system states, and its energy is considered to be energy of such states (for example, see [1]). It is assumed that environment influence results in the fact that system energy belongs to a small (but macroscopical) interval (E, E +ΔE). These makes it possible to consider a set of system states within the interval (E, E +ΔE), which are deemed to be of equal probability. It is considered, that this set represents a microcanonical ensemble of system states.

Apparently it is namely the environment influence that should ensure equal probability of such states both in the center of interval and on its ends. Such sophistication of all equilibrium systems' surrounding is not to understand. Besides the process of internal equilibrium attainment influenced by surrounding would critically depend on the system's surface quality and the environmental structure.

When canonical distribution is derived, a system S is considered as a part of a very big system U described by microcanonical distribution [1]. Interaction of system S with surrounding W (or addition to system U) is considered, on one hand, necessary for an exchange of energy, but, on the other hand, negligibly small – to make it possible to talk about certain quantum state $m$ of the system S with the energy $E_m$. In other words, S and W are considered to be practically unconnected and not influencing each other.

Traditional way of deriving of canonical destribution the energy of surrounding $E_u - E_m$ corresponds to the value of the energy of the system S equal to $E_m$. It means that $E_m$ is considered as the total energy of the system S equal to some quantum value, and the system S can appear in states with various values of $E_m$ exclusively thanks to an energy exchange with the surrounding.

To receive observed values of physical quantities by means of canonical distribution, it is necessary for the system S to appear repeatedly in all states *m* during measurement. It means that there should be extremely fast energy exchange with the surrounding, i.e. the interaction energy should be very high. However, in this case it makes no sense to say that the system S is reliably in this or that quantum state. Besides, the system, which is practically unconnected with the surrounding, will not take care about the state of the environment, and the assumption that the big system U is in equilibrium is not physically well-founded. On the whole we can say that the assumptions made for the traditional derivation of canonical distribution contain insurmountable internal contradictions.

Canonical distribution provides the system with possibility to be in any part of the spectrum with respective probability, but not only within narrow interval (E, E +ΔE) around the value of system's total energy. Whereas the system spectrum does not depend on total energy value, the spectral region with energy values close to total energy does not have any advantages. But microcanonical ensemble includes eigenstates of system, belonging only to the region, and we can not consider other states.

Now remember, that in quantum mechanics system states with definite value of energy may be set by wave functions which are a linear combination of system energy operator eigenfunctions:

$$\psi = \sum_m a_m(t) \psi_m. \qquad (1)$$

where

$$a_m(t) = a_m(0) \exp\left(-\frac{i}{\hbar} E_m t\right).$$

$|a_m(t)|^2$ defines the probability of the system to be in a quantum state with energy $E_m$. As $|a_m(t)|^2 = |a_m(0)|^2$, we see that the quantum mechanics does not allow the system to pass into a state with a set of $|a_m(t)|^2$ different from the initial.

Normalisation requirement gives us the following:

$$\sum_m |a_m(t)|^2 = 1 \qquad (2)$$

In compliance with (1) total system energy may be written as follows:

$$E = \sum_m |a_m(t)|^2 E_m. \qquad (3)$$

If a number of values of system energy is more than two, equations (2) and (3) have a great number of solutions regarding $|a_m(t)|^2$. This means that value of total energy $E$ corresponds to a great number of different system states which may not be received one from another as a result of temporal system evolution.

We consider separating only eigenstates with different energy lying within the certain value interval (E, E +ΔE). caused by surrounding influence from the whole states of system to be out of reason. Instead we consider the whole set of states with the given system energy $E$ as states of microcanonical ensemble. We can't imagine how the system surrounding can ensure transitions within this set resulting in equality of states probability. In our view an equality of aprioristic probabilities must have a physical substantiation connected with internal processes in the system. However, according to quantum mechanics at every instant the system can be in the single state corresponding to a certain value of its total energy and the given initial conditions. In quantum mechanics there is no mechanism of transitions between states with different initial conditions which could ensure equality of aprioristic probabilities.

Thus, we can explain equality of aprioristic probabilities neither by macrosystem surrounding influence nor by its evolution in accordance with quantum mechanics. To find a way out of situation transpired let us remember that quantum mechanics creation was a reply to classical physics' inability to explain phenomena occurring in the microworld. Quantum-mechanical calculations, where they are possible, i.e. in systems with small number of particles, result in the remarkable accord with the experiment. Accuracy wherewith quantum mechanics gives hydrogen atom spectrum can serve an example. However, we have no reason to asseverate that these equations describe behavior of systems with macroscopically large number of particles with absolute accuracy.

Experiments [ 2,3] were carried out under supervision of the author of this article, where Hamiltonian sign of an isolated spin system reversed with predetermined accuracy. As the general solution of the Schrodinger equation has the following form:

$$\psi(t) = e^{-i\frac{\hat{H}}{\hbar}t}\psi(0)$$

it is evident that Hamiltonian sign reversal is identical to time sign reversal and it must result in the system reversion into initial state. However, it turned out that the experiments [ 2,3] cannot be correctly described on the basis of reversible equations of quantum mechanics. Note that otherwise it would break the second law of thermodynamics.

There was an assumption made in papers [2,3] mentioning that quantum-mechanical probability does not cover all probabilistic nature of the microworld and the God plays dice not exactly the way prescribed by Schrodinger. It would appear reasonable that for macrosystem, its states with equal energy in some sense are equivalent and there are transitions between states with equal energy not following Schrodinger equation and resulting in equality of aprioristic probabilities of these states. Probability of such transitions, which is not manifested in the systems with small number of particles, is considered in the work [4] as the reason of irreversibility in macrosystem evolution and the ground for microcanonical ensemble application in statistical mechanics. This probability is called X-probability in the work [4]. Of course, X- probability is connected with the nature of the system, with its energy characteristics, such as energy of single particles and value of interaction between them.

Certain manifold within phase space corresponds to a set of system states with the definite value of energy in a classical case. We can introduce space of system state function expansion coefficients $a_m$. Certain manifold will correspond to the definite value of the system total energy in this space, as well as in phase space of classical system. Dots on this manifold correspond to the states of microcanonical ensemble. The existence of X-probability provides with equality of mean values of physical quantity by time of a measurement and by microcanonical ensemble. The mean by microcanonical ensemble value of $a_m$ corresponds with canonical distribution.

As an illustration let us look at microcanonical ensemble of a system, having just three energy levels. In this case equations (2) and (3) will lead to the following:

$$|a_2|^2 = \frac{E - E_3}{E_2 - E_3} + |a_1|^2 \frac{E_3 - E_1}{E_2 - E_3} \qquad (4)$$

$$|a_3|^2 = \frac{E_2 - E}{E_2 - E_3} + |a_1|^2 \frac{E_1 - E_2}{E_2 - E_3} \qquad (5)$$

Because $0 < |a_m|^2 < 1$, we find:

$$\frac{E_3 - E}{E_3 - E_1} > |a_1|^2 > \frac{E_2 - E}{E_3 - E_1}$$

$$\frac{E_2 - E}{E_2 - E_1} < |a_1|^2 < \frac{E_3 - E}{E_3 - E_1} \qquad (6)$$

$$0 < |a_1|^2 < 1,$$

Conditions (6) must be met simultaneously. This defines the limitations of variations of $|a_1|^2$. Every value of $|a_1|^2$ corresponds with specific value of $|a_2|^2$ and $|a_3|^2$, that is specific state of the system. By changing $|a_1|^2$ in the limits allowed by (6) we can find all totality of states. Considering all the states of totality equally probable, we have microcanonical ensemble for the system.

Considering, that system state is defined by the first power of $a_m$, we write down the expression for the mean $|a_1|^2$ by the ensemble as:

$$\langle |a_1|^2 \rangle_m = \frac{1}{\sqrt{a} - \sqrt{b}} \int_{\sqrt{b}}^{\sqrt{a}} x^2 \, dx \qquad (7)$$

Here a and b are maximum and minimum of $|a_1|^2$ respectively.

Let's, for example, $E = 0, E = 5, E = 8, E = 2$. In this case, from the conditions (6) we find: $\frac{3}{5} < |a_1|^2 < \frac{3}{4}$.

Assuming in (7) $a = \frac{3}{4}$, $b = \frac{3}{5}$, we derive: $(|a_1|^2)_m = 0.674$. Substituting this value of $(|a_1|^2)_m$ in formulas (4) and (5), we find : $(|a_2|^2)_m = 0.204$, $(|a_3|^2)_m = 0.123$.

Let us show, that canonical distribution

$$P_n = Z^{-1} e^{-\beta E_n}, \qquad (8)$$

$$Z = \sum_k e^{-\beta E_k},$$

describes the results of calculations with good precision.

We find reverse temperature β, from condition:

$$E = \sum_n P_n E_n.$$

(9)

In our case we have:

$$5Z^{-1} e^{-\beta 5} + 8Z^{-1} e^{-\beta 8} = 2, \qquad (10)$$

$$Z = 1 + e^{-\beta 5} + e^{-\beta 8}.$$

Value of β = 0.223 corresponds to the expression (10). Substituting this value in (8), we receive for the probabilities of the system to be in quantum states with energy $E_m$: $P_1 = 0.669, P_2 = 0.2192, P_3 = 0.1122$.

Let's suppose now the full energy is equal 3 with previous values of energy levels. In this case mean by microcanonical ensemble values of probabilities $|a_i|^2$ are equal to: $(|a_1|^2)_m = 0.508, (|a_2|^2)_m = 0.3111, (|a_3|^2)_m = 0.1805$.

The reverse temperature of the system decreases with energy increase and with E=3 equals to 0.1199. Distribution (8) in this case gives: $P_1 = 0.5175, P_2 = 0.2842, P_3 = 0.1983$.

Thus, the difference of probabilities $P_k$ and mean by microcanonical ensemble $(|a_k|^2)_m$ in presented cases is not greater than 10%. We can get similar results for different values of energy levels and full energy of a three level system. The reader can sure on one's own that with the growth of a number of system energy levels the proximity of $(|a_k|^2)_m$ and $P_k$ only grows.

In macrosystems canonical distribution well describes the results of experiments even with a little measurement time. This means, that because of X-probability, the system has time to be in many states, uniformly distributed throughout all totality of states with the given energy. With the decrease of a number of particles in a system, X-probability decreases. This can lead to the mismatch of a results of system parameters measurements and their values, calculated using microcanonical distribution, and we can estimate the value of X-probability.

The author hopes, that proposed concept will help for deeper understanding of the statistical mechanic foundations and will stimulate to further research the lows of microworld.